# Comparative Quantum Cosmology: Causality, Singularity, and Boundary Conditions

**Philip V. Fellman**
Southern New Hampshire University
Manchester, NH
Shirogitsune99@yahoo.com

**Jonathan Vos Post**
Computer Futures
Altadena, CA
Jvospost3@gmail.com

**Christine M. Carmichael**
Woodbury University
Burbank, CA
**christine.carmichael@woodbury.edu**

**Andrew Carmichael Post**
University of Southern California
Los Angeles, CA
andrewpost@gmail.com

# 1.0 Introduction

Stuart Kauffman and Lee Smolin provide us with an interesting introduction to some of the problems of modern cosmology in "A Possible Solution For The Problem Of Time In Quantum Cosmology":[1]

> Cosmology, which came into its own as a science only about thirty years ago, is concerned in part with pinning down the parameters of the universe: its expansion rate, the amount of its mass, the nature of its "dark matter." Cosmologists today are also speculating on more far-reaching questions, such as how the universe was created and how its structure was determined. While some cosmologists are speculating that the laws of physics might explain the origin of the universe, the origin of the laws themselves is a problem so unfathomable that it is rarely discussed. Might the principles of adaptive complexity be at work? Is there a way in which the universe may have organized itself? Does the "anthropic principle", the notion that the existence of intelligent observers like us is in some sense a factor in the universe's existence have any useful part to play in cosmology?

Sean Carroll has explored this concept from a slightly different angle at length in "Is Our Universe Natural?" Carroll argues that while the concept of "natural" is somewhat difficult to specify in a precise physical systems context, particularly with respect to the laws of nature which govern the formation of the universe, some of the problems associated with entropy levels and scale levels can be resolved through an evolutionary mechanics.[2] In particular, he proposes a Multiverse situated in ordinary DeSitter space with positive vacuum energy, where quantum fluctuations can lead to the creation of multiple baby universes (see Appendix I).

---

[1] Smolin approaches the problem in another way in his discussion of General Relativity in "Einstein's Legacy – Where are the Einsteinians", arguing "General relativity is the most radical and challenging of Einstein's discoveries—so much so that I believe the majority of physicists, even theoretical physicists, have yet to fully incorporate it into their thinking. The flashy stuff, like black holes, gravitational waves, the expanding universe, and the Big Bang are, it turns out, the easy parts of general relativity. The theory goes much deeper: It demands a radical change in how we think of space and time. All previous theories said that space and time have a fixed structure and that it is this structure that gives rise to the properties of things in the world, by giving every object a place and every event a time. In the transition from Aristotle to Newton to special relativity, that structure changed, but in each case the structure remained fixed. We and everything that we observe live in a space-time, with fixed and unchanging properties. That is the stage on which we play, but nothing we do or could do affects the structure of space and time themselves. General relativity is not about adding to those structures. It is not even about substituting those structures for a list of possible new structures. It rejects the whole idea that space and time are fixed at all. Instead, in general relativity the properties of space and time evolve dynamically in interaction with everything they contain. Furthermore, the essence of space and time now are just a set of relationships between events that take place in the history of the world. It is sufficient, it turns out, to speak only of two kinds of relationships: how events are related to each other causally (the order in which they unfold) and how many events are contained within a given interval of time, measured by a standard clock (how quickly they unfold relative to each other). Thus, in general relativity there is no fixed framework, no stage on which the world plays itself out. There is only an evolving network of relationships, making up the history of space, time, and matter. All the previous theories described space and time as fixed backgrounds on which things happen. The point of general relativity is that there is no background." http://www.logosjournal.com/issue_4.3/smolin.htm

[2] "What makes a situation "natural"? Ever since Newton, we have divided the description of physical systems into two parts: the configuration of the system, characterizing the particular state it is in at some specific time, and the dynamical laws governing its evolution. For either part of this description, we have an intuitive notion that certain possibilities are more robust than others. For configurations, the concept of entropy quantifies how likely a situation is. If we find a collection of gas molecules in a high-entropy state distributed uniformly in a box, we are not surprised, whereas if we find the molecules huddled in a low-entropy configuration in one corner of the box we imagine there must be some explanation. For dynamical laws, the concept of naturalness can be harder to quantify. As a rule of thumb, we expect dimensionless parameters in a theory (including ratios of dimensionful parameters such as mass scales) to be of order unity, not too large nor too small. Indeed, in the context of effective quantum field theories, the renormalization group gives us some justification for this notion of "naturalness". In field theory, the dynamics of the low-energy degrees of freedom fall into universality classes that do not depend on the detailed structure of physics at arbitrarily high energies. If an interaction becomes stronger at large distances, we expect it to be relevant at low energies, while interactions that become weaker are irrelevant; anything else would be deemed unnatural. If any system should be natural, it's the universe. Nevertheless, according to the criteria just described, the universe we observe seems dramatically unnatural. The entropy of the universe isn't nearly as large as it could be, although it is at least increasing; for some reason, the early universe was in a state of incredibly low entropy. And our fundamental theories of physics involve huge hierarchies between the energy scales3 characteristic of gravitation (the reduced Planck scale, $E_{pl} = 1/\sqrt{8\pi G} \sim 10^{27}$ electron volts), particle physics (the Fermi scale of the weak interactions, $E_F \sim 10^{11}$ eV, and the scale of quantum chromodynamics, $EQCD \sim 10^8$ eV), and the recently-discovered vacuum energy ($E_{vac} \sim 10^{-3}$ eV).4 Of course, it may simply be that the universe is what it is, and these are brute facts we have to live with. More optimistically, however, these apparently delicately-tuned features of our universe may be clues that can help guide us to a deeper understanding of the laws of nature.

Quantum cosmology is an inherently difficult field. Stephen Hawking captures some of these difficulties in a 1999 lecture:[3]

> Cosmology used to be regarded as a pseudo science, an area where wild speculation, was unconstrained by any reliable observations. We now have lots and lots of observational data, and a generally agreed picture of how the universe is evolving. But cosmology is still not a proper science, in the sense that as usually practiced, it has no predictive power. Our observations tell us the present state of the universe, and we can run the equations backward, to calculate what the universe was like at earlier times. But all that tells us is that the universe is as it is now, because it was as it was then. To go further, and be a real science, cosmology would have to predict how the universe should be. We could then test its predictions against observation, like in any other science.

While the observational data has increased over the past eight years, competing theories of cosmology and quantum cosmology have proliferated to embrace the data rather than reducing to a few simple views. In addition, while the available data has increased, most of our recent observations have not been of a determinative nature. In the language of John Platt (1964), we do not have the data necessary to perform a critical experiment which could exclude many of the currently popular theories from scientific consideration (for additional details, see Smolin, 2007).

Quantum cosmology is further complicated by the fact that in attempting to link quantum mechanics (including quantum electrodynamics and quantum chromodynamics) with general relativity, there are *deep* problems within quantum mechanics itself, which are not entirely settled questions within the physics community, particularly those questions which have to do with issues of non-locality. As Smolin puts it:

> I am convinced that quantum mechanics is not a final theory. I believe this because I have never encountered an interpretation of the present formulation of quantum mechanics that makes sense to me. I have studied most of them in depth and thought hard about them, and in the end I still can't make real sense of quantum theory as it stands. Among other issues, the measurement problem seems impossible to resolve without changing the physical theory.
>
> Quantum mechanics must then be an approximate description of a more fundamental physical theory. There must then be hidden variables, which are averaged over to derive the approximate, probabilistic description which is quantum theory. We know from the experimental falsifications of the Bell inequalities that any theory which agrees with quantum mechanics on a range of experiments where it has been checked must be non-local. Quantum mechanics is non-local, as are all proposals for replacing it with something that makes more sense. So any additional hidden variables must be non-local. But I believe we can say more. I believe that the hidden variables represent relationships between the particles we do see, which are hidden because they are non-local and connect widely separated particles. This fits in with another core belief of mine, which derives from general relativity, which is that the fundamental properties of physical entities are a set of relationships, which evolve dynamically. There are no intrinsic, non-relational properties, and there is no fixed background, such as Newtonian space and time, which exists just to give things properties.

In addition, there are specific complications which non-locality introduces into the permutation of structures proposed in various species of string theory. The most significant complication is that the cross products of Bell's inequality work only for systems of three or seven physical dimensions, a measurement

---

[3]Hawking, S. (1999) "Physics Colloquiums - Quantum Cosmology, M-theory and the Anthropic Principle (January '99)" http://www.hawking.org.uk/text/physics/quantum.html  Hawking further argues, "The task of making predictions in cosmology is made more difficult by the singularity theorems, that Roger Penrose and I proved. These showed that if General Relativity were correct, the universe would have begun with a singularity. Of course, we would expect classical General Relativity to break down near a singularity, when quantum gravitational effects have to be taken into account. So what the singularity theorems are really telling us, is that the universe had a quantum origin, and that we need a theory of quantum cosmology, if we are to predict the present state of the universe. A theory of quantum cosmology has three aspects. (See Appendix III)

which 11d-brane string theorists currently regard as a coincidence, but one which if not mere coincidence, might impose severe constraints on contemporary string theories.

## 2.0 Entropy

One of the deep puzzles in the evolution of the universe is the level of observed entropy. We draw here on Carroll's characterization. Drawing on Boltzmann, he argues that the universe in initial conditions must have had extremely low entropy. He extends this logic to the time evolution of entropy, noting that entropy remains rather low in the universe today.[4] However a consideration of the entropy of black holes complicates the discussion. Using the Beckenstein-Hawking calculation of the entropy of a black hole, Carroll demonstrates that the entropy of a single (massive) black hole at the center of just one galaxy is higher than the entropy of all the ordinary matter in the visible universe:

$$S_{BH} \sim 10^{89} \left( \frac{M_{BH}}{10^6 M_\odot} \right)^2$$

Further, he argues that there should be some $10^{10}$ of these black holes in the universe, with a total entropy on the order of $10^{99}$. While there are additional calculations which can expand this number, the key point here is that in the early universe, there were no black holes, so that entropy would have had to have been extremely low, since it would have only been the entropy of normal matter. Understanding these conditions is difficult, but Carroll, like others in the field, relates this condition to homogeneity and isotropy (see also Wald, 2005). He cautions, however, that this may be a misleading approach to understanding the origin of the thermodynamic arrow of time. The problem, from his viewpoint, is that while it is tempting to associate the thermodynamic arrow of time with inflation, the converse, can also be argued – that inflation never works to decrease entropy, and that this theory "posits that our observable universe originates in a small patch whose entropy is fantastically less than it could have been, so that in fact there is a hidden fine-tuning of initial conditions implicit in the inflationary scenario, which consequently (in the extreme version of this reasoning) doesn't explain anything at all."[5] (Carroll and Chen, 2004). Carroll and Chen, then devote the remainder of their paper to the effort of resolving this fundamental tension in the second law.

From their perspective, this enterprise is aimed at understanding how the thermodynamic arrow of time might arise from initial conditions of the universe. In this regard, they define the problem quite precisely, and in a way which is significantly different than, for example, the approach used by Hawking and Penrose:[6]

> To make any progress we need to have some understanding of what it means for an initial condition to be "natural." If an unknown principle of physics demands specific initial conditions (but not final ones), there is no problem to be solved; it may simply be that the universe began in a low-entropy initial state and has been evolving normally ever since. This might be the case, for example, if the universe were created "from nothing," such that the initial state was a priori different from the final state, or if Penrose's explicitly time-asymmetric Weyl Curvature Hypothesis were true. There is, of course, no way to rule out this possibility.

---

[4] Within our observable patch, the entropy in ordinary matter is of order $S_M(U) \sim 10^{88}$ (Carroll and Chen, 2004)

[5] As they further develop their argument, Carroll and Chen consider the reversibility critique of inflation, arguing (p. 13) "The reversibility critique is thus very simple: the measure on the space of conditions that anti-inflate in a contracting universe is a negligible fraction of that on the space that collapse inhomogeneously (or even relatively homogeneously, but without anti-inflation); but the ratio of these measures is the same as the ratio of the measure on the space of inflationary initial conditions to that on the space of all conditions that evolve into a universe macroscopically like ours. Therefore, if our initial conditions are truly chosen randomly, inflationary conditions are much less likely to be chosen than some other conditions that evolve into our universe (which are still a tiny fraction of all possible initial conditions). In other words, nothing is really gained in terms of naturalness by invoking an early period of inflation; as far as random initial conditions go, it requires much less fine-tuning to simply put the universe in some state that can evolve into our present conditions. The idea that our current universe is more likely to be randomly chosen than a small, smooth patch dominated by a large potential energy seems intuitively nonsensical. Our current universe is large, complicated, and filled with particles, while the proto-inflationary patch is tiny, simple, and practically empty. But our intuition has been trained in circumstances where the volume, or particle number, or total energy is typically kept fixed, and none of these is true in the context of quantum field theory coupled to gravity." Note how their final argument also ties to issues of quantum gravity raised by Smolin.

[6] Ibid.

Nevertheless, it would be more satisfying if we could somehow understand our apparently low-entropy initial condition as an outcome of dynamical evolution from a generic state. In particular, we wish to avoid any explicit violation of time-reversal symmetry in the specification of the initial condition. The very word "initial," of course, entails a violation of this symmetry; we should therefore apply analogous standards to the final conditions of the universe. Price has proposed a "double-standard principle" – anything that is purportedly natural about an initial condition for the universe is equally natural when applied to a final condition. The conventional big-bang model clearly invokes unnatural initial conditions, as we would not expect the end state of a recollapsing universe to arrange itself into an extremely homogeneous configuration…

Carroll and Chen then go on to illustrate the problems of developing a consistent explanation which embodies all of these characteristics. After exhausting the exposition of possible models with low entropy states at both the beginning and end of the universe, Carroll and Chen suggest what they describe as a "loophole" in the double-standard argument by considering a situation where there is high entropy at both the beginning and the end of the universe.[7] As they so elegantly state their problem (p. 7):

We are left with the following conundrum: we would like to explain our currently observed universe as arising from natural initial conditions, but natural means high-entropy, and high-entropy implies equilibrium configurations with occasional fluctuations, but not ones sufficient to explain our observed universe. However, there is one loophole in this reasoning, namely the assumption that there is such a thing as a state of maximal entropy. If the universe truly has an infinite number of degrees of freedom, and can evolve in a direction of increasing entropy from any specified state, then an explanation for the observed arrow of time arises more naturally.

## 3.0 Entropy and Inflation

Their treatment is to analogize the problem of entropy in the initial conditions of the universe to a particle moving in a potential that rolls off to infinity without having any minima. They illustrate this as follows (p.8):

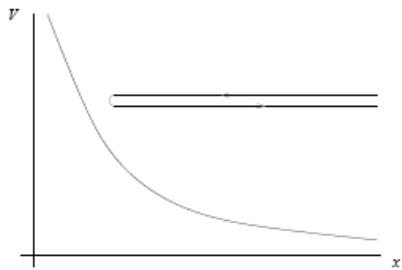

Figure 1: Motion of a particle in a potential that rolls to infinity without ever reaching a minimum. All trajectories in this potential come in from infinity, reach a turning point, and return to infinity.

---

[7] Along the lines suggested by Smolin, Carroll and Chen (2004) argue at one point, "One way to understand the concept of naturalness would be to imagine that we had a measure on the space of microstates describing the universe, and an appropriate way of coarse-graining into macroscopically indistinguishable states. The entropy is then simply the logarithm of the measure on any collection of macroscopically indistinguishable microstates, and natural initial conditions are those with large entropy. Unfortunately, we do not have a reliable understanding of entropy in the context of quantum gravity; the fundamental degrees of freedom themselves remain unclear. Even without a quantitative understanding at the level of statistical mechanics, however, there are certain essential features that entropy must have, which will be sufficient to highlight problems with the conventional picture." (p. 6)

In this context, any local state, except the turning point can be viewed as either coming from or going to a state with higher values. In approaching entropy in this fashion, they then define "the past" as simply the direction of time in which entropy is *locally* decreasing. (p. 7) Of course this explanation depends upon a universe with infinitely many degrees of freedom, however, this caveat, then provides a mechanism for "natural" time evolution of entropic states. And, in at least one sense, this is still the background dynamics for Carroll and Chen's explanation, as they subsequently argue, "Of course, it is certainly not sufficient to imagine that the entropy of the universe is unbounded, although it seems to be necessary. We also need to understand why the process of entropy creation would create regions of spacetime resembling our observable universe." (p. 8) And it is in this context that they develop their inflationary model.

After considering and dismissing a variety of thermodynamic theories, Carroll and Chen argue (p. 19):

> The examples considered in this section provide anecdotal evidence for a straightforward claim: in a theory with gravity (and vanishing vacuum energy), the closest thing to a maximal-entropy, thermal-equilibrium state is flat empty space. Another way to reach the same conclusion is to simply take any configuration defined on a spacelike hypersurface, and realize that we can increase the entropy by taking the same set of excitations (whether in matter fields or gravitational waves) and spread them apart by increasing the scale factor, thereby increasing the allowed phase space for each particle. This expansion doesn't violate any conservation laws of the system, so there is no obstacle to configurations eventually increasing their entropy in this fashion. Consequently, there is no reason to expect randomly-chosen or generic conditions to feature large curvatures and Planck-scale fluctuations. According to everything we know about gravity, large curvatures are entropically disfavored, tending to ultimately smooth themselves out under ordinary evolution. This is a direct consequence of the ability of a curved spacetime to evolve towards perpetually higher entropy by having the universe expand, unlike gas trapped in a box. From this point of view, it should not be considered surprising that we live in a relatively cold, low-curvature universe; the surprise is rather that there is any observable matter at all, much less evolution from an extremely hot and dense Big Bang.

Having put the question in this fashion , they acknowledge that this makes developing an inflationary model not less, but vastly more difficult. However, despite these difficulties, the core of their explanation resides in a rather compact model of inflation based upon spontaneous inflation from cold DeSitter space. The introduce this model with the argument that empty spacetime corresponds to DeSitter rather than Minkowski spacetime and thus it has a Gibbons-Hawking temperature defined as (p. 20):

$$T_{\mathrm{dS}} = \frac{H}{2\pi} \sim \frac{M_{\mathrm{vac}}^2}{M_{\mathrm{pl}}}$$

It is this temperature which gives rise to thermal fluctuations and their explanation follows with a description of how fluctuations in the appropriate field $\phi$ can lead to the spontaneous onset of inflation, which "can then continue forever as in the standard story of eternal inflation" (pp. 20-21)[8]

---

[8] Carroll and Chen explain, This idea is not new; Garriga and Vilenkin, for example, have proposed that thermal fluctuations can induce tunneling from a true de Sitter vacuum to a false vacuum at higher energies, thus inducing spontaneous inflation. There is also a body of literature that addresses the creation of inflating universes via quantum tunneling, either "from nothing," at finite temperature, or from a small patch of false vacuum. In our discussion is that we examine the case of an harmonic oscillator potential without any false vacua; in such a potential we can simply fluctuate up without any tunneling. The resulting period of inflation can then end via conventional slow-roll, which is more phenomenologically acceptable than tunneling from a false vacuum (as in "old inflation"). Thus, the emptying-out of the universe under typical evolution of a generic state can actually provide appropriate initial conditions for the onset of inflation, which then leads to regions that look like our universe. We should emphasize that our calculation ignores many important subtleties, most obviously the back-reaction of the metric on the fluctuating scalar field. Nevertheless, our goal is to be as conservative as possible, given the limited state of our current understanding of quantum gravity. In particular, it is quite possible that a similar tunneling into inflation may occur even in a Minkowski background. In our calculation we simply discard the vacuum fluctuations that are present in Minkowski, and examine only the additional contributions from the nonzero de Sitter temperature. We believe that the resulting number (which is fantastically small) provides a sensible minimum value for the probability to fluctuate up into inflation. The true answer may very well be bigger; for our purposes, the numerical result is much less important

## 4.0 Time

In 2003, Peter Lynds published a controversial paper, "Time and Classical and Quantum Mechanics: Indeterminacy vs. Discontinuity" in Foundations of Physics Letters.[9] Lynds' theory does away with the notion of "instants" of time[10], relegates the "flow of time" to the psychological domain.[11] While we have commented elsewhere on the implications of Lynds' theory for mathematical modeling[12] it might save a bit of time simply to borrow Wikipedia's summary of this paper:

> Lynds' work involves the subject of time. The main conclusion of his paper is that there is a necessary trade off of all precise physical magnitudes at a time, for their continuity over time. More specifically, that there is not an instant in time underlying an object's motion, and as its position is constantly changing over time, and as such, never determined, it also does not have a determined relative position. Lynds posits that this is also the correct resolution of Zeno's paradoxes, with the paradoxes arising because people have wrongly assumed that an object in motion has a determined relative position at any given instant in time, thus rendering the body's motion static and frozen at that instant and enabling the impossible situation of the paradoxes to be derived. A further implication of this conclusion is that if there is no such thing as determined relative position, velocity, acceleration, momentum, mass, energy and all other physical magnitudes, cannot be precisely determined at any time either. Other implications of Lynds' work are that time does not flow, that in relation to indeterminacy in precise physical magnitude, the micro and macroscopic are inextricably linked and both a part of the same parcel, rather than just a case of the former underlying and contributing to the latter, that Chronons, proposed atoms of time, cannot exist, that it does not appear necessary for time to emerge or congeal from the big bang, and that Stephen Hawking's theory of Imaginary time would appear to be meaningless, as it is the relative order of events that is relevant, not the direction of time itself, because time does not go in any direction. Consequently, it is meaningless for the order of a sequence of events to be imaginary, or at right angles, relative to another order of events.

---

than the simple fact that the background is unstable to the onset of spontaneous inflation. Clearly this issue deserves further study." (pp. 20-21)

[9] Foundations of Physics Letters. (Vol. 16, Issue 4, 2003).

[10] Ibid., "…although it is generally not realized, in all cases a time value indicates an interval of time, rather than a precise static instant in time at which the relative position of a body in relative motion or a specific physical magnitude would theoretically be precisely determined. For example, if two separate events are measured to take place at either 1 hour or 10.00 seconds, these two values indicate the events occurred during the time intervals of 1 and 1.99999… hours and 10.00 and 10.0099999… seconds, respectively. If a time measurement is made smaller and more accurate, the value comes closer to an accurate measure of an interval in time and the corresponding parameter and boundary of a specific physical magnitudes potential measurement during that interval, whether it be relative position, momentum, energy or other. Regardless of how small and accurate the value is made however, it cannot indicate a precise static instant in time at which a value would theoretically be precisely determined, because there is not a precise static instant in time underlying a dynamical physical process. If there were, all physical continuity, including motion and variation in all physical magnitudes would not be possible, as they would be frozen static at that precise instant, remaining that way. Subsequently, at no time is the relative position of a body in relative motion or a physical magnitude precisely determined, whether during a measured time interval, however small, or at a precise static instant in time, as at no time is it not constantly changing and undetermined. Thus, it is exactly due to there not being a precise static instant in time underlying a dynamical physical process, and the relative motion of body in relative motion or a physical magnitude not being precisely determined at any time, that motion and variation in physical magnitudes is possible: there is a necessary trade off of all precisely determined physical values at a time, for their continuity through time." (p. 1)

[11] In "Subjective Perception of Time and a Progressive Present Moment: The Neurobiological Key to Unlocking Consciousness", http://cogprints.org/3125/ Lynds provides a compelling and novel explanation for why the human nervous system perceives time as *flowing*. While it has been argued by some critics of his physical theory that this is a philosophical position rather than hard physics, Lynds' neurobiology paper is perhaps the first advance since the advent of the parallellistic hypothesis in 20th century philosophy of science to explain the subjective perception of time in terms of the behavior of physical systems.

[12] See "The Implications of Peter Lynds 'Time and Classical and Quantum Mechanics: Indeterminacy vs. Discontinuity' for Mathematical Modeling", Proceedings of the North American Association for Computation in the Social and Organizational Sciences, Carnegie Mellon University, http://www.casos.cs.cmu.edu/events/conferences/2004/conference_papers.php and http://www.casos.cs.cmu.edu/events/conferences/2004/2004_proceedings/Lynds.v2doc.doc

One can see from the above summary that this radical reformulation of the concept of time is bound to have significant cosmological implications. We discussed some of these implications in a brief paper in 2004, "Time and Classical and Quantum Mechanics and the Arrow of Time".[13] We began with a discussion of John Gribbin's analysis, "Quantum Time Waits for No Cosmos", and Gribbin's argument against the mechanics of time reversibility as an explanation for the origin of the universe, where he cites Raymond LaFlamme:[14]

> The intriguing notion that time might run backwards when the Universe collapses has run into difficulties. Raymond LaFlamme, of the Los Alamos National Laboratory in New Mexico, has carried out a new calculation which suggests that the Universe cannot start out uniform, go through a cycle of expansion and collapse, and end up in a uniform state. It could start out disordered, expand, and then collapse back into disorder. But, since the COBE data show that our Universe was born in a smooth and uniform state, this symmetric possibility cannot be applied to the real Universe.

The concept of time reversibility, while apparently quite straightforward in many cases, seems never to be without considerable difficulty in cosmology, and indeed, explaining the mechanics of time reversibility and its relationship to Einstein's cosmological constant is one of the major enterprises of quantum cosmology (Sorkin, 2007). The terns of the debate expressed by Gribbin above, have been extended by Wald (2005) who acknowledges Carroll and Chen's work, discussed earlier in this paper, but who argues that at some point an anthropic principle introduces a circularity into the causal logic. Sorkin covers much of the distance needed for a Carroll-Chen type counterargument in his 2007 paper, "Is the cosmological "constant" a nonlocal quantum residue of discreteness of the causal set type?", however a complete discussion of Sorkin's model is beyond the scope of our present exploration.[15]

## 5.0 The Arrow of Time

One set of problems with the thermodynamic arrow of time for the very early universe (and this is partially addressed by Sorkin) is that in addition to the problem of possible inhomogeneities in the early universe there is still a lack of consensus or unequivocal evidence on the invariance of fundamental physical constants during the early history of the universe. Various authors have recently suggested that the speed of light may have been greater during the earliest period of the universe's formation (Murphy, Webb and Flambaum, 2002).

In our 2004 review of Lynds work on how we might better understand the thermodynamic arrow of time we also raised two moderately troubling complexity issues which at present remain largely unanswered. The first is the problem of heteroskedastic time behavior in the early universe. This question is not unconnected to the question of changes in the values of fundamental physical constants in the early universe. In most models of early universe formation a smooth or linear flow of time is assumed. However, it is possible to imagine inflationary models where the expansion of time dimension or the time-like dimensions of a higher order manifold inflate in a heteroskedastic fashion. To the extent that the thermodynamic arrow of time is invoked as an element of cosmological explanation, It would need to be

---

able to explain the dynamical evolution of the universe, not just as we know it today, but at those particularly difficult to characterize beginning and end points of the system. The difficulty with heteroskedastic time distributions is that they may or may not allow recovery of the standard Boltzmann expression. At a deeper level, it is likely that in characterizing the development of the early universe, one may have to incorporate a significant number of non-commuting quantum operators. Further, in this context, our knowledge of the early universe is both substantively incomplete, because we lack any system of measurement for the first three hundred thousand years of time evolution of the system (i.e. the period prior to the formation of baryons and photons) and very likely theoretically incomplete as well. We can compare the problem to one of discrete time series evolutions with low dimensionality and discrete combinatorics. For example, the random order of a shuffled deck of cards can eventually be repeated because the dimensionality of the system is low. As dynamical systems take on higher orders of dimensionality their asymptotes become ill-defined (in at least one sense, this is the objection raised by Gribbin and LaFlamme).[16]

Another problem is what Freeman Dyson characterizes as the "struggle between order and entropy in the big crunch. As the universe approaches infinity and the average density approaches zero, temperature does not approach zero, and thus the nature of the struggle between order and entropy may actually be characterized by very different time evolutions that those with which we are familiar. In addition, there is the "Maxwell's Demon" family of arguments. This is a systems dynamic which is particularly relevant to complexity science. The problem here is that there may be emergent phenomena at the end of the life of the universe which causes the system's time evolution to then behave in unexpected ways. In some sense this is logical trap lurking behind statistical reasoning. Under normal conditions, the descriptive and inferential statistical conjecture that the near future will look like the recent past (or more boldly that fundamental physical constants are perfect invariants) is entirely reasonable. However, in the face of emergent phenomena, this assumption may not hold. Indeed, this problem is at the center of much of the debate over "relic" radiation and arguments over the age of the universe.

Yet another problem, which may also encompass emergent behavior, has to do with symmetry breaking. The universe has undergone several phase transitions by symmetry breaking. As a result, additional forces have emerged at each of these transitions. First gravity separated out of other forces, and it is for that reason that we can expect gravity wave detectors to probe more deeply into the early history of the universe than any other technology. To return to the emergent properties argument, we cannot definitively rule out (by means of present theory and observations) the possibility that at some future time (presumably near the end of the system's time evolution) that some fifth force will separate out from the known four forces.[17] At the classical level, time reversibility and a thermodynamic arrow of time is no longer problematical, but at the quantum level, and at the cosmological level, the concept remains murky at best.

## 6.0 Lynds' Conjecture

Peter Lynds has developed an alternative cosmology, or an alternative foundation for cosmology which flows in part from his treatment of time. He introduces his approach by stating:[18]

> Based on the conjecture that rather than the second law of thermodynamics inevitably be breached as matter approaches a big crunch or a black hole singularity, the order of events should reverse, a model of the universe that resolves a number of longstanding problems and paradoxes in cosmology is presented. A universe that has no beginning (and no need for one), no ending, but yet is finite, is without singularities, precludes time travel, in which events are neither determined by initial or final conditions, and problems such as why the universe has a low entropy past, or conditions at the big bang appear to be so special, require no causal explanation, is the result. This model also has some profound philosophical implications.

---

[16] A substantial amount of work on non-extensive statistical mechanics has been done by Tsallis, et al. most recently (2007) "Nonergodicity and Central Limit Behavior for Long-range Hamiltonians" http://arxiv.org/PS_cache/arxiv/pdf/0706/0706.4021v3.pdf

[17] Admittedly, this is a significant part of the epistemological argument put forth by Carroll and Carroll and Chen in the "natural" universe conjecture. Implicit in their theory is the idea that if cosmos formation is a "natural" phenomena, rather than the "unnatural" situation suggested by the differences in scale values for fundamental constants, then an additional, emergent force would be "unnatural".

[18] Lynds, P. (2006) "On a finite universe with no beginning or end", http://arxiv.org/ftp/physics/papers/0612/0612053.pdf

The model arises in part as a consequence of Lynds' unique treatment of time, and his ability to present a scientific framework which dispenses with the conventional notion of "instants" and a concomitant "flow" of these instants of time. He develops his cosmology based on the conjecture that in a "big crunch", at precisely the moment where the second law of thermodynamics would necessarily be breached in order to preserve symmetrical event structure, and just before the universe reaches a singularity, instead of breaching the second law of thermodynamics, the order of events would be reversed and universal expansion would begin without a singularity actually having been reached. In Lynds words:[19]

> The natural question then became, what would happen if the second law of thermodynamics were breached? People such as Hawking [8, 11] and Gold [9,10] had assumed that all physical processes would go into reverse. In other words, they had assumed that events would take place in the direction in which entropy was decreasing, rather than increasing as we observe today. Furthermore, they had assumed that entropy would decrease in the direction in which the universe contracted towards a big crunch (in their case, towards what we call the big bang). But if the second law correctly holds, on a large scale, entropy should still always increase. Indeed, what marks it out so much from the other laws of physics in the first place, is that it is asymmetric - it is not reversible. If all of the laws of physics, with the exception of the second law of thermodynamics, are time symmetric and can equally be reversed, it became apparent that if faced with a situation where entropy might be forced to decrease rather than increase, rather than actually doing so, the order of events should simply reverse, so that the order in which they took place would still be in the direction in which entropy was increasing. The second law would continue to hold, events would remain continuous, and no other law of physics would be contravened.

Hence, in Lynds' model, any events which would have taken place in a situation where entropy was decreasing would experience a reversal of the time ordering of events, and in the subsequent expansion of the universe, "events would immediately take up at where the big crunch singularity would have been had events not reversed, and in this direction, no singularity would be encountered. The universe would then expand from where the big crunch singularity would have been had events not reversed (i.e. the big crunch reversed), and with events going in this direction, entropy would still be increasing, no singularity would be encountered, and no law of physics would be contravened. They would all still hold." (p. 6)

Lynds' argument necessarily bounds this reversal in the ordering of events to a very small region, and quite shortly thereafter, normal processes of inflation, including increasing entropy resume. Both the physical and the philosophical implications of this position are profound. On the philosophical side, Lynds has introduced a new concept, not only of the ultimate origin of the universe, but also a complex redefinition of "past" and "future":

> At this point, it becomes apparent that this would not only lead things back to the big bang, but it would actually cause it. The universe would then expand, cool, and eventually our solar system would take shape. It would also mean that this would be the exact repeat of the universe we live in now. Something further becomes evident, however, and it is perhaps the most important (and will probably be the most misunderstood and puzzled over) feature of this model. If one asks the question, what caused the big bang?, the answer here is the big crunch. This is strange enough. But is the big crunch in the past or the future of the big bang? It could equally be said to be either. Likewise, is the big bang in the past or future of the big crunch? Again, it could equally be said to be either. The differentiation between past and future becomes completely meaningless. Moreover, one is now faced with a universe that has neither a beginning nor end in time, but yet is also finite and needs no beginning. The finite vs. infinite paradox of Kant completely disappears.

---

[19] Ibid.

Although if viewed from our normal conception of past and future (where we make a differentiation), the universe would repeat over and over an infinite number of times, and could also be said to have done so in the past. Crucially, however, if one thinks about what is actually happening in respect to time, no universe is in the future or past of another one. It is exactly the same version, once, and it is non-cyclic. If so desired, one might also picture the situation as an infinite number of the same universe repeating at exactly the same time. But again, if properly taking into account what is happening in respect to time, in actuality, there is no infinite number of universes. It is one and the same.

As previously indicated, this conjecture represents another radical and novel interpretation of time. However, one of the most interesting features of Lynds' conjecture is that it actually meets the two primary criteria of Hawking's M-Theory, (a) weak singularity (in Lynds' case the singularity is there, but it is, in some sense, outside the light cone and outside the observable event horizon) and (b) no boundary condition (albeit, not in the precise fashion that Hawking interprets the no boundary condition restriction).[20] Admittedly, the model is in some ways profoundly counter-intuitive, but that is largely because in a very curious way, even when treating subjects in both relativity and quantum mechanics, we have a tendency to either overtly or covertly introduce Newtonian notions of time. Some of this is addressed in Smolin's critiques of general relativity and quantum mechanics.[21] Lynds himself addresses a potential source of difficulty in the section of his article entitled "Potential Criticisms": (p.9)

An obvious criticism for the model seems to raise itself. It implies that the universe can somehow anticipate future or past events in exact detail, and then play them over at will. At first glance, this just seems too far-fetched. How could it possibly *know*? With a little more thought, however, one recognizes that such a contention would assume that there actually was a differentiation between past and future events in the universe. With this model, it is clear there would not be. Events could neither be said to be in the future or past of one another; they would just be. Moreover, as there is nothing to make one time (as indicated by a clock) any more special than another, there is no present moment gradually unfolding; all different events and times share equal reality (in respect to time, none except for the interval used as the reference). Although physical continuity (i.e. the capability for events to be continuous), and as such, the capability for motion and of clocks and rulers to represent intervals, would stop them from all happening at once (and to happen at all), all events and times in the universe would already be mapped out. As such, as long as it still obeyed all of its own physical laws, the universe would be free and able to play any order of events it wished. Please note that this timeless picture of reality is actually the same as that provided by relativity and the "block" universe model [26], the formalized view of space-time resulting from the lack of a "preferred" present moment in Einstein's relativity theories, in which all times and events in the universe - past, present and future - are all mapped out together, fixed, and share equal status.

---

[20] It is important to note that when the clock restarts at the big bang, the universe is not in the future or past of another one. In a sense, it is time itself that restarts (although, again, nothing in fact actually "restarts"), so there is no past or future universe. Because of this, no conservation laws are violated. It is also important to note that it is simply just the order of events that reverse - something that would be immediate. Time does not begin "flowing" backwards to the big bang, nor does anything travel into the future or past of anything, including time and some imagined "present moment" [22, 23]. Indeed, this model contains another interesting consequence. As there is no differentiation between past and future in it, and, strictly speaking, no event could ever be said to be in the future or past of another one, it would appear to provide a clear reason as to why time travel is not possible. In relation to future and past, there is clearly nothing there to travel into. Physically speaking, the same can be said for travel through an interval of time, a flow of it, as well as space-time [22, 23]. (p. 8)

[21] In "Three Roads to Quantum Gravity", Smolin argues that the fundamental flaw in relativity is that it fails to incorporate the effect of the observer on observed phenomena and that quantum mechanics, while achieving the former has a tendency to treat quantum-mechanical events as occurring in traditional, Newtonian spacetime. He then argues that the unification of the partial completeness of these two new physical paradigms will be required to develop an adequate theory of quantum gravity. A quantum cosmology is likewise implicit in such a unification. Lynds provides some interesting clues to this unification insofar as puts time on all scales on a firm Einsteinian footing. In fact, one might answer Smolin's provocative essay title "Where are the Einsteinians?" with the retort "In New Zealand".

Lynds' model contains a number of additional features, including some novel treatments of Kaon decay, black holes and "white holes", all of which are successfully incorporated in his model. While it is beyond the scope of the present paper to discuss these details, they deserve mention as indicators of the level of sophistication in what some might initially imagine to be a naïve interpretation of quantum cosmology.

## 7.0 Conclusion

In the foregoing review paper, we have briefly examined several competing theories of quantum, cosmology, with particular attention to the arguments proposed by Smolin, Carroll, Hawking and their collaborators. We have also directed attention to Peter Lynds' work, which in some ways may provide a unifying basis for theories of quantum cosmology. A deep problem in quantum cosmology is that many of the concepts which it uses as foundational building blocks become ill defined when applied to the early universe, and in this context we lack measurement methods for much of the first three hundred thousand years or so of the early history of the universe. To the extent that quantum cosmologies represent an attempt at a dynamical explanation of the history of system, this lack of definition is profoundly troubling. While Lynds conjecture does not provide a complete solution to this problem, it suggests both a useful heuristic and a novel *gestalt* for understanding these problems.

## Appendix I: Carroll's Pre-Inflation Multiverse Model[22]

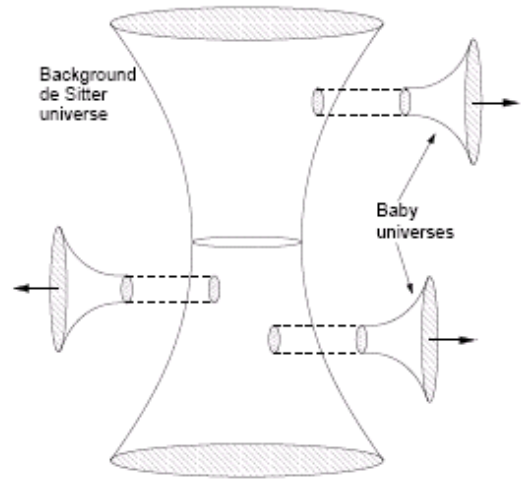

Figure 2: A possible spacetime diagram for the universe on ultra-large scales. A natural state for a universe with a positive vacuum energy is empty de Sitter space. In the presence of an appropriate scalar field, quantum fluctuations in such a background can lead to the nucleation of baby universes. Each baby universe is created in a proto-inflationary state, which then expands and reheats into a universe like that we observe.

In a related paper, "Spontaneous Inflation and the Origin of the Arrow of Time", Carroll and Chen (2004)[23] argue that "spontaneous eternal inflation can provide a natural explanation for the thermodynamic arrow of time…In the absence of inflation, we argue that systems coupled to gravity usually evolve asymptotically to the vacuum, which is the only natural state in a thermodynamic sense.  In the presence of a small vacuum energy and an appropriate inflation field, the DeSitter vacuum is unstable to the spontaneous onset of inflation at higher energy scale.  Starting from DeSitter, inflation can increase the total entropy of the universe without bound, creating universes similar to ours in the process.  An important consequence of this picture is that inflation occurs asymptotically both forwards and backwards in time, implying a universe that is (statistically) time symmetric on ultra-large scales.

---

# Appendix II:  Topologically stable braids in loop quantum gravity and spin foam models.[24]

There is an old dream that matter is topological excitations of the geometry of space-time.  Recently it was discovered that this is realized in the context of models of quantum geometry based on spin networks, such as those used in loop quantum gravity and spin foam models[1, 2, 3]. Although the rough idea that topological features of space-time geometry might be interpreted as particles is very old, two questions delayed implementation of the idea till recently. First, how do we identify an independent excitation or degree of freedom in a background independent quantum theory, where the semi-classical approximation is expected to be unreliable? Second, why should such excitations be chiral, as any excitations that give rise to the low mass observed fermions must be?

Recently a new approach to the first question was proposed in [1, 2], which was to apply the notion of noise free subsystem from quantum information theory to quantum gravity, and use it to characterize an emergent elementary particle. The key point is that noiseless subsystems arise when a splitting of the Hilbert space of the whole system (in this case the whole quantum geometry) into system and environment reveals emergent symmetries that can protect subsystems from decohering as a result of inherently noisy interactions with the environment. The proposal raised the question of whether there were models of dynamical quantum geometry in which this procedure reproduced at least some of the symmetries of the observed elementary particles.

This led to an answer to the second question. In [3] it was shown that this was the case in a simple model of the dynamics of quantum geometry. The result of that paper incorporated prior results of [4] where a preon model was coded into a game of braided triplets of ribbons. The needed ribbon graphs are present in models related to loop quantum gravity when the cosmological constant is non-zero[5, 6]; and the three ribbon braids needed are indeed the simplest systems in that model bound by the conservation of topological quantum numbers. Strikingly, the results of [3] make contact with both structures from topological quantum computing and now classic results of knot theory from the 1980s.  In both chirality plays a key role because braids are chiral and topological invariants associated with braided ribbons are able to detect chirality and code chiral conservation  laws.

---

[24] Taken from Smolin, L. and Wan, Y., (2007) "Propagation and interaction of chiral states in quantum gravity" http://arxiv.org/PS_cache/arxiv/pdf/0710/0710.1548v1.pdf

# Appendix III: Stephen Hawking's Three Pillars of Quantum Cosmology[25]

The task of making predictions in cosmology is made more difficult by the singularity theorems, that Roger Penrose and I proved. These showed that if General Relativity were correct, the universe would have begun with a singularity. Of course, we would expect classical General Relativity to break down near a singularity, when quantum gravitational effects have to be taken into account. So what the singularity theorems are really telling us, is that the universe had a quantum origin, and that we need a theory of quantum cosmology, if we are to predict the present state of the universe. A theory of quantum cosmology has three aspects.

The first, is the local theory that the fields in space-time obey. The second, is the boundary conditions for the fields. And I shall argue that the anthropic principle, is an essential third element. As far as the local theory is concerned, the best, and indeed the only consistent way we know, to describe gravitational forces, is curved space-time. And the theory has to incorporate super symmetry, because otherwise the uncancelled vacuum energies of all the modes would curl space-time into a tiny ball. These two requirements, seemed to point to supergravity theories, at least until 1985. But then the fashion changed suddenly. People declared that supergravity was only a low energy effective theory, because the higher loops probably diverged, though no one was brave, or fool hardy enough to calculate an eight-loop diagram. Instead, the fundamental theory was claimed to be super strings, which were thought to be finite to all loops. But it was discovered that strings were just one member, of a wider class of extended objects, called p-branes. It seems natural to adopt the principle of p-brane democracy. All p-branes are created equal. Yet for p greater than one, the quantum theory of p-branes, diverges for higher loops.

I think we should interpret these loop divergences, not as a breakdown of the supergravity theories, but as a breakdown of naive perturbation theory. In gauge theories, we know that perturbation theory breaks down at strong coupling. In quantum gravity, the role of the gauge coupling, is played by the energy of a particle. In a quantum loop one integrates over… So one would expect perturbation theory, to break down. In gauge theories, one can often use duality, to relate a strongly coupled theory, where perturbation theory is bad, to a weakly coupled one, in which it is good. The situation seems to be similar in gravity, with the relation between ultra violet and infra red cut-offs, in the anti de Sitter, conformal field theory, correspondence. I shall therefore not worry about the higher loop divergences, and use eleven-dimensional supergravity, as the local description of the universe. This also goes under the name of M theory, for those that rubbished supergravity in the 80s, and don't want to admit it was basically correct. In fact, as I shall show, it seems the origin of the universe, is in a regime in which first order perturbation theory, is a good approximation. The second pillar of quantum cosmology, are boundary conditions for the local theory. There are three candidates, the pre big bang scenario, the tunneling hypothesis, and the no boundary proposal.

The pre big bang scenario claims that the boundary condition, is some vacuum state in the infinite past. But if this vacuum state develops into the universe we have now, it must be unstable. And if it is unstable, it wouldn't be a vacuum state, and it wouldn't have lasted an infinite time before becoming unstable. The quantum-tunneling hypothesis, is not actually a boundary condition on the space-time fields, but on the Wheeler Dewitt equation. However, the Wheeler Dewitt equation, acts on the infinite dimensional space of all fields on a hyper surface, and is not well defined. Also, the 3+1, or 10+1 split, is putting apart that which God, or Einstein, has joined together. In my opinion therefore, neither the pre bang scenario, nor the quantum-tunneling hypothesis, are viable. To determine what happens in the universe, we need to specify the boundary conditions, on the field configurations, that are summed over in the path integral. One natural choice, would be metrics that are asymptotically Euclidean, or asymptotically anti de Sitter. These would be the relevant boundary conditions for scattering calculations, where one sends particles in from infinity, and measures what comes back out. However, they are not the appropriate boundary conditions for cosmology.

We have no reason to believe the universe is asymptotically Euclidean, or anti de Sitter. Even if it were, we are not concerned about measurements at infinity, but in a finite region in the interior. For such measurements, there will be a contribution from metrics that are compact, without boundary. The action of a compact metric is given by integrating the Lagrangian. Thus its contribution to the path integral is well defined. By contrast, the action of a non-compact or singular metric involves a surface term at infinity, or at the singularity. One can add an arbitrary quantity to this surface term. It therefore seems more natural to adopt what Jim Hartle and I called the no boundary proposal. The quantum state of the universe is defined by a Euclidean path integral over compact metrics. In other words, the boundary condition of the universe is that it has no boundary.

There are compact Reechi flat metrics of any dimension, many with high dimensional modulie spaces. Thus eleven-dimensional supergravity, or M theory, admits a very large number of solutions and compactifications. There may be some principle that we haven't yet thought of, that restricts the possible models to a small sub class, but it seems unlikely. Thus I believe that we have to invoke the Anthropic Principle. Many physicists dislike the Anthropic Principle. They feel it is messy and vague, it can be used to explain almost anything, and it has little predictive power. I sympathize with these feelings, but the Anthropic Principle seems essential in quantum cosmology. Otherwise, why should we live in a four dimensional world, and not eleven, or some other number of dimensions. The anthropic answer is that two spatial dimensions, are not enough for complicated structures, like intelligent beings. On the other hand, four or more spatial dimensions would mean that gravitational and electric forces would fall off faster than the inverse square law. In this situation, planets would not have stable orbits around their star, nor electrons have stable orbits around the nucleus of an atom. Thus intelligent life, at least as we know it, could exist only in four dimensions. I very much doubt we will find a non anthropic explanation…